\def\breakon{\end{multicols}\widetext\vspace{-.7cm}
\noindent\rule{.49\linewidth}{.3mm}\rule{.3mm}{.5cm}\vspace{0.0cm}}
\def\breakoff{\vspace{-.25cm}
\noindent
\rule{.50\linewidth}{.0mm}\rule[-.47cm]{.3mm}{.5cm}\rule{.489\linewidth}{.3mm}
\vspace{-.55cm}
\begin{multicols}{2}
\narrowtext\noindent}
\def\ie{{\it i.e.\ }}
\documentstyle[aps,multicol]{revtex}
\begin{document}
\title{Transition from Abelian to non-Abelian FQHE states}
\author{D.C.\ Cabra$^{1,2}$, A.\ Lopez$^3$ and G.L.\ Rossini$^1$}
\address{$^1$Departamento de F\'{\i}sica, Universidad Nacional de la Plata,
C.C.\ 67, (1900) La Plata, Argentina.\\ $^2$ Facultad de
Ingenier\'{\i}a, Universidad Nacional de Lomas de Zamora,\\ Cno.\
de Cintura y Juan XXIII, (1832), Lomas de Zamora, Argentina.\\$^3$
Centro At{\'o}mico Bariloche, 8400 S.\ C.\ de Bariloche, R{\'\i}o
Negro, Argentina}
\bigskip

\maketitle

\begin{abstract}
We study the transition  from the Abelian multi-component
$(3,3,1)$ quantum Hall state to the non-Abelian one component
Pfaffian state in bilayer two dimensional electron systems. We
show that tunneling between layers can induce this transition. At
the transition points part of the degrees  of freedom that
describe the $(3,3,1)$ state disappear from the spectrum, and the
system is correctly  described by the Pfaffian state, with
quasi-particles that  satisfy non-Abelian statistics. The
mechanism described in this work provides for a physical
Hamiltonian interpretation of the algebraic projection from the
$(3,3,1)$ to the Pfaffian state that has been discussed in the
literature.

\end{abstract}
\bigskip

\begin{multicols}{2}
\narrowtext

Even denominator states in double layer two dimensional electron
systems (2DES) have been observed experimentally \cite{eis} and
are theoretically quite well understood \cite{PDS}. The two 2DES
are separated by a potential barrier that, if high and thick
enough, will inhibit both Coulomb interactions and tunneling
between layers. If the barrier is made thinner, Coulomb
interactions will become important even if tunneling is still
suppressed. The relevant parameter to measure this effect is the
ratio $d/l_0$ where $d$ is the interlayer separation and $l_0$ is
the magnetic length. In real samples neither Coulomb interactions
nor tunneling can be completely neglected. Therefore a very rich
phase diagram can be constructed with Coulomb interlayer
interaction (or alternatively the distance $d$) on one axis and
the tunneling amplitude  on the other.

We will concentrate here on  systems in which a quantized Hall
plateau exists at total filling fraction $\nu=1/2$. The phase
diagram for these systems was first discussed  by
Halperin\cite{halperin}. He assumed that the actual spin of the
electrons was polarized in the direction of the external field,
and that the two layers were completely equivalent. A possible
experimental realization for this system is a single wide quantum
well in which the self consistent Coulomb potential creates a
barrier in the middle of the well with maxima in the electron
density at the two edges. Halperin suggested \cite{halperin} that
for an intermediate range of distances $d$ and vanishing
tunneling, the so called $(3,3,1)$ state should be a stable phase
for the system.

The $(3,3,1)$ state is a correlated bilayer state which is
basically stabilized by Coulomb interactions. It was shown
\cite{yoshioka} that if the layer separation is large enough the
state collapses into decoupled layers due to the fact that interlayer Coulomb
interactions become negligible. The variational wave function
describing this state, proposed by Halperin \cite{h}  in the
context of spinful systems has the form \breakon
\begin{equation}
\Psi_{331}= \prod_{i<j} (z_i-z_j)^3 \prod_{i<j} (w_i-w_j)^3\prod_{i,j}(z_i-w_j)
 e^{-{\frac 14} \sum (|z_i|^2+\vert w_i\vert^2)}\ ,
\label{eq:wf331}
\end{equation}
\breakoff where $z_i$ and $w_i$ are the coordinates of the
electrons in each plane. The first two factors represent the
correlations within each layer, and the last one corresponds to
the intralayer correlations. The $\nu=1/2$ state was observed
experimentally and it was checked numerically that its properties
are indeed well described by the $(3,3,1)$ wave function
(\ref{eq:wf331}) \cite{PDS}.

It was also conjectured \cite{halperin} that a transition to a
Pfaffian state should occur, within the range of distances $d$ for
which the $(3,3,1)$ state is stable at vanishing tunneling, when
the tunneling amplitude is made large enough.
The Pfaffian state, proposed by Moore and Read \cite{mr}, is a
candidate for a fractional quantum Hall state at $\nu=1/2$ in
single layer systems, or in general for $\nu=1/q$ where $q$ is an
even number (were we working with bosons with strong repulsive interactions, $q$ would
be an odd integer \cite{sch}). Its
variational wave function is given by
\begin{equation}
\Psi_{Pf}= {\rm Pfaff}({\frac {1}{z_i-z_j}}) \prod_{i<j} (z_i-z_j)^q
e^{-{\frac 14} \sum |z|^2} \ ,
\label{eq:wfPf}
\end{equation}
where the Pfaffian is defined for a $2N \times 2N$ antisymmetric matrix
whose elements are $M_{ij}$ by
\begin{equation}
{\rm Pfaff} (M_{ij})= \frac {1}{2^N N!} \sum_{\sigma \varepsilon
S_{2N}} {\rm sgn} (\sigma) \prod_{k=1}^N
M_{\sigma(2k-1),\sigma(2k)}
\label{eq:Pf}
\end{equation}
or as the square root of the determinant of $M$. It was shown
\cite{mr} that this wave function  arises from applying Wick's
theorem to real fermion fields, or as the real space BCS wave
function for pairing of spinless fermions.

The $\nu=1/2$ states were extensively studied in experiments
\cite{suen,suen2} in a wide single quantum well sample, varying the
well width and sheet density. It was concluded that the state
observed was the $(3,3,1)$, \ie the Pfaffian state did not show up
within the range of tunneling amplitude and thickness scanned in
the experiments. The authors argued nevertheless that it should
still appear in the phase diagram for larger tunneling.

Our goal is to explore the above mentioned transition between the
$(3,3,1)$ and the Pfaffian states. Therefore, we consider a system
in which the interlayer separation $d$ is kept fixed,
while the tunneling amplitude between layers can be changed arbitrarily.
In other words, we will be looking at Halperin's phase diagram for a
given value of the interlayer separation, such that if the tunneling
amplitude vanishes, the $(3,3,1)$ state is the stable phase of the system.

We start with the usual chiral boson approach for the edge
theory of the $(3,3,1)$ state (see {\it e.g.} \cite{wen}), that was
recently reviewed in \cite{naud} with the inclusion of
tunneling between layers. We further include a
chemical potential term for the electrons. In this case the
original theory, written in terms of two chiral bosons (a $c=2$
central charge Conformal Field Theory (CFT)), can be mapped into
an effective theory with one chiral boson and two Majorana
fermions. We then study the phase diagram as a function of electron
tunneling  $\lambda$ and chemical potential $\mu$.

As we have already mentioned, given that the spacing between layers
is kept fixed at a value such that both pahses are stable,
the Pfaffian state could describe a
double layer sample in the limit in which the tunneling amplitude
between the layers is large enough so as the two species of
electrons of the $(3,3,1)$ state become indistinguishable.
 Since
the edge theory for the Pfaffian state can be described by a
$c=3/2$ CFT \cite{mr}, the question is then how does this process
occur physically, \ie how does the $(3,3,1)$ CFT with $c=2$
evolve to the Pfaffian CFT with $c=3/2$. Related to this, it has been shown
\cite{sch,tod,wpt} that the $(3,3,1)$ edge theory can be seen as
the enveloping theory for the non Abelian Pfaffian state. Indeed,
there is an algebraic procedure by which the two elementary
quasi-holes of the $(3,3,1)$ state merge into one in the Pfaffian
state by getting rid of an Ising CFT factor from the original edge
theory. However, an explicit mechanism implementing physically
this procedure is, to our knowledge, still lacking. In this letter
we address this issue and show that electron tunneling between
layers is capable of implementing this projection. More precisely,
when the tunneling amplitude and/or the chemical potential
increase, there is a critical line in the $(\lambda , \mu)$ plane
at which one of the degrees of freedom that describes the original
theory disappears. We furthermore show that the remaining degrees
of freedom acquire the quantum numbers of the elementary
excitations for the Pfaffian state and non-Abelian statistics
emerges.

The edge theory for the $(3,3,1)$ state is described by the Hamiltonian \cite{wen}
\begin{equation}
H= {\frac {1}{4\pi}}\int dx  V_{ij} :\partial_x u_i \partial_x
u_j:~,
\label{eq:ho}
\end{equation}
where colons denote standard normal ordering.
Here $x$ is the coordinate along the edge, the $u_i$ are chiral  bosonic
fields whose compactification radius is $1$, and $V_{ij}$ is a
symmetric matrix whose coefficients depend on the confining potential
and the interparticle interactions at the edge,
\begin{equation}
V= \left(
\begin{array}{cc}
v & g \\
g & v
\end{array}
\right)~.
\label{eq:vij}
\end{equation}
The commutation relations for the bosonic fields are
\begin{equation}
[u_i(x,t),u_j(x',t)] = i \pi K_{ij} {\rm sgn}(x-x') ~,
\label{eq:cr}
\end{equation}
where K is a symmetric matrix which characterizes the
topological properties of the system
\begin{equation}
K= \left(
\begin{array}{cc}
3 & 1 \\
1 & 3
\end{array}
\right) ~.
\label{eq:kij}
\end{equation}

There exists an orthogonal transformation that diagonalizes $V$ and $K$
simultaneously, after which the  Hamiltonian eq.\ (\ref{eq:ho}) simply reads
\begin{equation}
H= {\frac {1}{4\pi}}\int dx[v_c :( \partial_x \phi_c)^2: + v_n :( \partial_x
\phi_n)^2:] ~,
\label{eq:hon}
\end{equation}
where $v_c=4(v+g)$ and $v_n=2(v-g)$. Notice that the condition $\det V
> 0$ must hold in order that both modes have the same chirality.
$\phi_c$ and $\phi_n$ refer to charged and neutral modes
respectively, which are chiral bosons
with standard commutation relations
\begin{equation}
[\phi_i(x,t),\phi_j(x',t)] = i \pi \delta_{ij} {\rm sgn}(x-x') ~.
\label{eq:crn}
\end{equation}

The electron operators can be written in this basis
as follows
\begin{eqnarray}
\psi_{e1}&\propto& :e^{i(-\sqrt2 \phi_c +\phi_n)}: \nonumber \\
\psi_{e2}&\propto& :e^{i(-\sqrt2 \phi_c -\phi_n)}:
\label{eq:el12}
\end{eqnarray}
while the quasi-particle operators are
\begin{eqnarray}
\psi_{qp1}&\propto& :e^{-i({\frac {1}{\sqrt8}}
\phi_c +{\frac {1}{2}}\phi_n)}: \nonumber \\
\psi_{qp2}&\propto& :e^{-i({\frac {1}{\sqrt8}}
\phi_c - {\frac {1}{2}}\phi_n)}:~.
\label{eq:qp12}
\end{eqnarray}

In reference \cite{naud} the authors considered the problem of
adding uniform electron tunneling to the edge theory. Here we will
study the same problem adding also a chemical potential for the
electrons. Therefore we add to the Hamiltonian the following
perturbation terms
\begin{eqnarray}
H'= -&&\mu_0 \int dx [:\psi_{e1}^\dagger \psi_{e1}+\psi_{e2}^\dagger
\psi_{e2}:] \nonumber \\ + && \lambda_0 \int dx [:\psi_{e1}^\dagger
\psi_{e2}+ \psi_{e2}^\dagger
\psi_{e1}:] ~.
\label{eq:hp}
\end{eqnarray}
Using the bosonic representation for the electron operators we can
write
\begin{eqnarray}
:\psi_{e1}^\dagger \psi_{e1}+\psi_{e2}^\dagger \psi_{e2}: &\propto&
(i 2\sqrt{2} a_0  \partial_x \phi_c  - a_0^2 :(\partial_x \phi_n)^2:)
\nonumber \\
: \psi_{e1}^\dagger \psi_{e2}+ \psi_{e2}^\dagger
\psi_{e1}: &\propto& : e^{-i2\phi_n} +
e^{i2\phi_n}: ~,
 \label{eq:bo}
\end{eqnarray}
where $a_0$ is the UV cut-off.
In terms of these bosons the total  Hamiltonian can be decoupled
into charged ($H_c$)  and neutral ($H_n$)  sectors given by
\begin{eqnarray}
 H_c  &=& \int dx [{\frac {1}{4\pi}} v_c
:( \partial_x \phi_c)^2:  - \mu_c
:\partial_x \phi_c:] \nonumber \\
H_n &=& \int dx  {\frac {1}{4\pi}}(v_n + \mu_n ):(
\partial_x \phi_n)^2:\nonumber \\
&-&  \int dx ~ \lambda  :(e^{-i2\phi_n} +
e^{i2\phi_n}):~, \nonumber \\ \label{eq:hbos}
\end{eqnarray}
where $\mu_c,\mu_n\propto \mu_0$ and $\lambda \propto \lambda_0$.

The properties of the charged sector are not changed by the
perturbation since the new term is linear in derivatives and can
be absorbed by a shift in the bare Hamiltonian.

As for the neutral mode, it proves useful to decompose it (through conformal
embedding) in terms of two chiral Majorana fermions  \cite{naud}

\begin{equation}
:e^{-i\phi_n}:\ \propto ( \chi_1 +i \chi_2) ~.
\label{eq:Majoranas}
\end{equation}
The Hamiltonian then reads

\breakon
\begin{equation}
H_n  =- {\frac {1}{2}}
\int dx \left(i( v_n +\mu_n +\lambda_{eff}) :\chi_1 \partial_x \chi_1: +
i( v_n +\mu_n -\lambda_{eff}) :\chi_2 \partial_x \chi_2: \right)~,
\label{eq:hmaj}
\end{equation}
\breakoff
where $\lambda_{eff}\propto \lambda$ \cite{comment}.

We see that the two chiral Majorana fermions behave as free
fields, but acquire different velocities which are determined by
the bare velocity of the neutral boson $v_n$, the tunneling
amplitude $\lambda_{eff}$ and the chemical potential $\mu_n$.
Moreover, each Majorana sector describes a (chiral) Ising CFT.

It is clear now that, assuming that the perturbative treatment of
the interaction Hamiltonian (\ref{eq:hp}) is valid, there are two
lines in the $(\lambda_{eff},\mu_n )$ plane, given by $\mu_n =
-(v_n  \pm \lambda_{eff})$, on which one of the Majorana
velocities vanishes. Though this observation is immediate from
eq.\ (\ref{eq:hmaj}), the study of the emerging state is
non-trivial and constitutes the main result in the present work.
The key observation is that when one of these two conditions is
satisfied, the corresponding Ising sector disappears from the
spectrum and, as we shall see, the remaining degrees of freedom
describe the physics of the Pfaffian state. In fact, the
Hamiltonian density for the zero-velocity mode vanishes, therefore
its energy-momentum tensor and hence its central charge vanish.
In this way, the central charge of the original system (the
$(3,3,1)$ state) decreases by  $1/2$. The remaining system is
described by one chiral boson and one Majorana fermion with total
central charge $c_{eff} = 3/2$, which is the correct value for
describing the Pfaffian state.

To make sure that the projection procedure drives the system to
the Pfaffian state, we now show how the electron and
quasi-particle operators (\ref{eq:el12}), (\ref{eq:qp12}) in the
$(3,3,1)$ phase come to describe the corresponding operators in
this new phase. To this end, we rewrite the original electron and
quasi-particle operators in terms of the charged boson and the
Ising primary fields (the Majorana fermions $\chi_a$, the spin
(order) operators $\sigma_a$ and their duals (disorder) $\mu_a$,
where $a=1,2$ labels the two Ising sectors). Therefore the
electron operators for the $(3,3,1)$ phase in eq.\ (\ref{eq:el12})
can be written as
\begin{eqnarray}
\psi_{e1} \propto \  :e^{-i\sqrt2 \phi_c} (\chi_1 +i \chi_2):
\nonumber
\\ \psi_{e2} \propto \  :e^{-i\sqrt2 \phi_c} (\chi_1 -i \chi_2):~.
\label{eq:el12n}
\end{eqnarray}
The neutral components of the quasi-particle operators can be
combined and represented in terms of the order and disorder fields
$\sigma_a$ and $\mu_a$ as
\begin{eqnarray}
:e^{i\phi_n/2}+e^{-i\phi_n/2}:\  \propto \sigma_1\otimes
\sigma_2,\nonumber\\
:e^{i\phi_n/2}-e^{-i\phi_n/2}:\  \propto  \mu_1\otimes
\mu_2 ~.
\label{qpn}
\end{eqnarray}
This identification has been proven in reference \cite{Di Francesco et al.} by a
careful analysis of operator product expansions on both sides.
Then the quasi-particle operators can be written as
\begin{eqnarray}
\psi_{qp1}+\psi_{qp2} \propto \
:e^{-i\frac{1}{\sqrt8} \phi_c}\sigma_1\otimes \sigma_2:\ ,
\nonumber\\
\psi_{qp1}-\psi_{qp2} \propto \  :e^{-i\frac{1}{\sqrt8} \phi_c}\mu_1\otimes
\mu_2: ~.
\label{qp}
\end{eqnarray}

The vanishing of the energy-momentum tensor for one of the two
Ising sectors at the critical line implements a coset
construction. The essence of the coset is to project a sector out
from the physical Hilbert space. In the case at hand the projected
subspace corresponds to one of the Ising sectors (say $a=2$) of
the $(3,3,1)$ theory \cite{sch}.

It should be stressed at this point that the coset projection
appears in a natural way within this context. Previous treatments
advocating the coset mechanism for projecting out an Ising sector
were performed without any connection to a Hamiltonian
description.

An important question that remains to be answered is how this
projection acts on the quasi-particle and electron operators. This
projection can be seen as if all
primaries in the projected sector become trivial (they have
vanishing conformal weights).
More precisely, the quasi-particle operators (\ref{eq:qp12})
degenerate into a single quasi-particle operator

\breakon
\begin{equation}
\left.
\begin{array}{l}
\psi_{qp1}\propto \ :e^{-i\frac{1}{\sqrt8} \phi_c}(\sigma_1\otimes
\sigma_2 + \mu_1\otimes \mu_2): \\ \psi_{qp2}\propto \
:e^{-i\frac{1}{\sqrt8} \phi_c}(\sigma_1\otimes \sigma_2-
\mu_1\otimes \mu_2):
\end{array}
\right\}
\rightarrow
\psi_{qp}^{\rm Pfaff}\propto \ :e^{-i\frac{1}{\sqrt8} \phi_c} \sigma_1
:\
\label{eq:Pfaffqp}
\end{equation}
\breakoff
(there are indeed two possible dual descriptions in terms of
$\sigma$ or its dual $\mu$) describing quasi-particle excitations
over the Pfaffian ground state. They have charge $e/4$ and, more importantly,
exhibit non-Abelian statistics.

Besides, the two original electron operators are projected onto

\breakon
\begin{equation}
\left.
\begin{array}{l}
\psi_{e1}\propto \ :e^{-i\sqrt2 \phi_c}: (\chi_1\otimes 1_2 +i 1_1 \otimes
\chi_2) \\ \psi_{e2}\propto \ :e^{-i\sqrt2 \phi_c}: (\chi_1 \otimes 1_2 -i
1_1 \otimes \chi_2)
\end{array}
\right\} \rightarrow \psi_e^{\rm Pfaff} \pm :e^{-i\sqrt2
\phi_c}:\ \propto \ :e^{-i\sqrt2 \phi_c}: \chi_1 \pm :e^{-i\sqrt2 \phi_c}:\ ,
\label{eq:electronprojection}
\end{equation}
\breakoff
that is the Pfaffian electron operator plus a four quasi-particle bound state
$:e^{-i\sqrt2 \phi_c}:$ ({\it c.f.} eq.\ (\ref{eq:Pfaffqp})).

Once the electron and quasi-particle operators at the edge are
known,  the (bulk) wave functions for both the ground state and
excited states can be constructed following  \cite{mr}, by
computing suitable correlation functions of those operators. In
this way one recovers the expression in eq.\ (\ref{eq:wfPf}) for
the Pfaffian ground state. The computation of the wave function
for four quasi-holes over the ground state shows that the
quasi-particle statistics is non-Abelian.

It is worth mentioning that non-Abelian statistics arises in this
context due to the fact that one of the order-disorder fields
becomes trivial. The corresponding computation with the full
(non projected) quasi-particle operators gives the correct Abelian
statistics in the $(3,3,1)$ phase.

In summary, we have shown that if we add to the edge theory for
the $(3,3,1)$ state tunneling and  chemical potential terms, there
exists a critical line where part of the degrees of freedom that
describe the system becomes unphysical and disappears from the
spectrum. This is precisely the line where the characteristic
properties  of the $(3,3,1)$ state are lost. In turn,  at these
points of the parameter space, the electron and quasi-particle
operators of the $(3,3,1)$ state can be mapped into the
corresponding operators of the Pfaffian state, and the
statistics of quasi-particles becomes non-Abelian. The  mechanism
described in this work provides for a physical interpretation of
the algebraic projection from the $(3,3,1)$ to the Pfaffian state
that has been discussed in the literature \cite{sch,tod,wpt}. The
question that remains to be answered is whether the Pfaffian state
corresponds to a stable phase, \ie if the system remains in this
state beyond the critical lines. An alternative treatment to the
one presented here is eventually needed to resolve this issue
within the framework of the edge theories. According to the phase
diagram found by comparing the Pfaffian state bulk wave function
with the real space BCS wave function for pairing of spinless
fermions  \cite{gr,ho}, the system should be in a Pfaffian phase
somewhere beyond those lines.

\noindent {\it Acknowledgements: } We acknowledge useful
discussions with P.\ Degiovanni, E.\ Fradkin, A.\ Honecker, A.\ Lugo, E.F.\
Moreno, P.\ Pujol and I.\ Todorov. This work is
partially supported by CONICET, ANPCyT through grants No.\
03-02249 and 03-03924 (AL), and Fundaci\'on Antorchas (Argentina)
through grants No.\ A-13622/1-18, A-13622/1-106 and A-13740/1-64.

\end{multicols}


\begin{references}
\bibitem{comment} Even though eq.\ (\ref{eq:hmaj}) has been already derived
in reference \cite{naud} in the case of $\mu_n =0$, we remark here
that the physical consequences of the vanishing of the velocity of
one of the Majorana fermions has not been discussed
 by their authors.
\end{references}
\end{document}